# Insulator and Electrode Materials Marginally Influence Carbonized Layer Conductivity in Metalized-Film Capacitors

Vitaly V. Chaban[1] and Nadezhda A. Andreeva[2]

(1) Yerevan State University, Yerevan, 0025, Armenia. E-mail: vvchaban@gmail.com.
(2) Peter the Great St. Petersburg Polytechnic University, Saint Petersburg, Russia.

**Abstract**. Capacitor self-healing is a generalized term to describe physical and chemical processes restoring the functionalities of a dielectric capacitor after an electrical breakdown. The efficacy of self-healing depends on the elemental composition of a metalized-film capacitor. We report atomistic simulations of self-healing from a chemical perspective proving the impossibility of tuning the electrical conductivity of the soot by finding an interplay of various polymers and electrodes. All investigated soot samples turn out to possess carbon-rich semiconducting skeletons with numerous unsaturated C-C covalent bonds. They exhibit electrical conductivities of the same order of magnitude, irrespective of initial chemical compositions and properties of the chosen insulating polymers. Upon reporting the new results, we discuss less evident approaches to diminish the soot conductivity. We conclude that the quality of capacitor self-healing can be assessed by counting gaseous by-products of electrical breakdown or evaluating the volume of the solid-state semiconducting counterpart.

**Keywords:** self-healing; dielectric capacitor; metalized-film capacitor; electrical breakdown; insulating polymer.

**TOC Graphic**

Semiconducting soot emerges in a capacitor as a result of electrical breakdown.

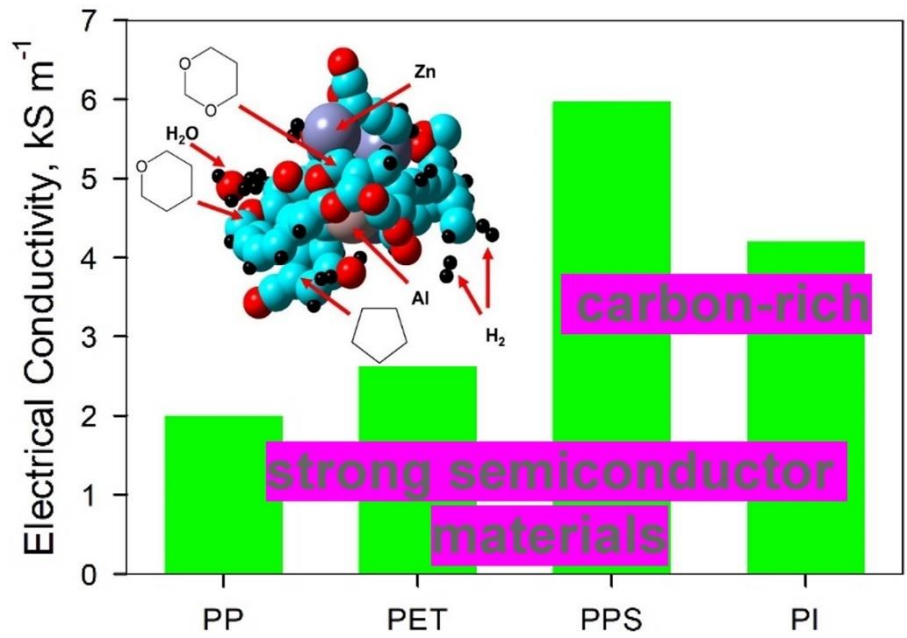

**Research Highlights**

The electrical breakdown produces adverse semiconducting soot.

The conductivity of the soot is weakly sensitive to the capacitor design.

The quality of self-healing depends on the amount of gaseous by-products.

**Introduction**

Self-healing in metalized film dielectric capacitors represents a spontaneous physical and chemical phenomenon taking place after the electrical breakdown.[1-2] As implied by the term, self-healing compensates for the adverse structural alterations. Self-healing is driven by the inherent thermodynamics of physical and chemical processes inside the device and does not involve any external control. However, the quality of self-healing does depend on the design of the capacitor and the elemental composition of its constituents.[2] As soon as electrical engineers possess relevant information, they can develop metalized-film capacitors not only with a due resilience to electrical breakdown but also with a high potential for self-healing.[3-6]

Meticulous experimental and theoretical investigation of self-healing is urged to develop dielectric capacitors with higher capacitance, lower self-discharge, and longer lifespan. Whereas this sort of research is traditionally pursued by physicists, it contains an extensive chemical component.[1,7] For instance, the electrical breakdown initiates numerous chemical reactions. Specifically, the electrodes and the dielectric materials first undergo thermal decomposition at subplasmic temperatures and then assemble into novel structures.[7] The solid phase called soot exhibits semiconducting properties. In turn, the gaseous products generate internal pressures inside the capacitor. The control over these processes and their optimization necessitates an engineer to acquire in-depth knowledge of materials sciences, physicochemical methods of research, and electrical phenomena.[1,7]

Computer simulations originating from quantum chemistry and potential energy landscape sampling[1] can be applied to describe the processes of electrical breakdown and further self-healing in dielectric capacitors. The thermally induced chemical destruction of the insulators and electrodes can be mimicked followed by the self-assembly of the soot samples and gas expulsion. Since the breakdown-related chemical reactions take place at high and very high temperatures, the role of the reaction activation barriers turns out to be marginal. The identities of the products are determined by thermochemical reasons, i.e., the most thermodynamically stable molecular and supramolecular structures emerge. By employing appropriate potential energy minimization algorithms, it is possible to predict and characterize the cornerstone chemical processes occurring upon the electrical breakdown.

The production of improved dielectric capacitors depends on the progress in polymer chemistry dramatically.[8-11] Polypropylene (PP)[8] is nowadays considered to be the most suitable dielectric material for metalized-film dielectric capacitors with zinc-aluminum-alloy electrodes.[12] The insulating film is stretched during the time of production to enhance exploitation properties. Yet, the principal limitation remains the inability to increase the operating temperature above 100 °C. In turn, polyethylene terephthalate (PET)[10] offers a relatively high dielectric constant and high dielectric strength.[12] While PET remains somewhat more kinetically stable around 100 °C compared to PP, its dielectric constant increases at elevated temperatures and frequencies. Some additional dielectric materials may appear useful to boost the thermal stabilities of films and manufacture lighter devices. In this context, polyimide (PI), polyphenylene sulfide (PPS), and polyethylene naphthalate are presently in focus of academic research efforts.[13] For the first time, we hereby employ mathematical atomistically precise modeling to unravel the mechanism and characterize the quality of self-healing in the case of PI and PPS.

In the reported simulations, we constructed the capacitor electrodes out of the Zn/Al alloys, in which the molar fraction of zinc amounts to 50 and 75%. Aluminum enhances the adhesion of the electrodes to the insulator films because of its electrophilic coupling with the dielectric. In turn, zinc serves as a core component of the electrode. It improves the electrode stability and costs modestly. As a compromise between capacitor lifespan and electrode atmospheric stability, the aluminum content in the Zn/Al alloys has been recommended not to exceed 10 wt%.[14] However, in simulations, we had to deliberately use significantly enhanced aluminum percentages because of the computational costs and, hence, the limited sizes of the model boxes. The goal was to have at least two Al atoms per system in order to detect possible Al…Al coupling cases in the potentially semiconducting soot. Such information may be of essential importance to understanding the trends in macroscopic soot samples.

There are three goals of the present atomistic simulations. First, we aim to chemically characterize the soot structures emerging after the electrical breakdown and unravel the role of the original elemental composition of a capacitor. Second, we aim to understand how the metal atoms previously belonging to the electrode rearrange and interact with the non-metal atoms previously belonging to the insulator film. Third, we aim to establish a correlation between the overall elemental compositions of a metalized film to the precise electrical conductivity of the semiconducting soot forming after the breakdown event.

**Methods and Methodology**

The kinetic energy injection method was used to let the system explore its potential energy landscape before applying the local geometry optimization to arrive at the stationary point. The performance of the method is based on periodically injecting random momenta into the system to assist it in overcoming potential energy barriers and freely traveling throughout the phase space. Three hundred perturbation cycles were sampled to obtain the corresponding number of non-unique stationary points. The entire composite methodology for this type of simulation[15] has been extensively described and validated elsewhere.[15-18] The numerical parameters employed in the present work are as follows: $10^{-6}$ hartree for the self-consistent-field algorithm convergence; $3\times10^{-2}$ hartree bohr$^{-1}$ for the maximum atomic force; $1\times10^{-2}$ hartree bohr$^{-1}$ for the root-mean-squared atomic force; 0.1 ps for the duration of sampling during one cycle; 0.0 ps and 300 K for the Berendsen thermostat relaxation constant[19] and reference temperature; 0.0001 ps for the Verlet time step of atomic nuclei; 5,000 K for the periodically perturbing kinetic energy. The PM7 Hamiltonian[20-22] was used to compute immediate atomic forces, instead of the force field models.

Plane-wave Kohn-Sham density functional theory (KSDFT) calculations with pure PBEPBE exchange-correlation functional[23] were used to re-optimize the low-energy geometries obtained from PM7 calculations, compute the electronic structure, and derive the electrical conductivities of the soot samples. Pure DFT specifically refers to functionals that rely solely on the electron density and its derivatives to calculate the system's energy. This contrasts with hybrid functionals, which incorporate a portion of exact exchange from Hartree-Fock's theory. Pure DFT methods are generally more computationally efficient than hybrid functionals, making them suitable for larger systems. Yet, pure DFT methods struggle to accurately describe systems with strong static correlation or dispersion interactions. Therefore, the VDWD correction[24] was applied to realistically represent van der Waals forces in KSDFT calculations. The number of k-points for all compositions was set to 27 to sample the Brillouin space. The plane-wave cut-off energy was set to 73 Ry. These parameters were obtained from the benchmarking calculations as exemplified in our previous works on capacitor soot samples.[1]

In-house programs were used to simulate minimum point search.[17,25-26] Quantum Espresso (version 6.0)[27] was used to optimize the geometries of the periodic systems, optimize the cell vectors, and derive band gaps and electrical conductivities. OpenMopac (https://github.com/openmopac/mopac; version 22) was used to compute PM7 formation enthalpies[20-21,28] whenever needed. The energy landscape exploration was conducted with the in-house software GMSEARCH (version 240203). VMD (version 1.9.3) was used to visualize molecular structures and molecular trajectories, producing detailed atomistic images.[29]

**Results and Discussion**

The simulated capacitor compositions containing both electrode and polymer atoms are enumerated in Table 1. Apart from considering four insulating films, we revealed the role of the elemental composition of the electrode. Depending on the insulating film, the metal atoms (Zn and Al) of the electrodes may acquire different forms in the soot, interacting with one another covalently and non-covalently.

Table 1. Fundamental parameters describing simulated chemical compositions.

| # | Composition | # nuclei | # electrons | Abbreviation |
|---|---|---|---|---|
| 1 | 2 Zn + 2 Al + [$C_3H_6$]$_{10}$ | 94 | 326 | 2 Zn + 2 Al + PP |
| 2 | 2 Zn + 2 Al + [$C_{10}H_8O_4$]$_5$ | 114 | 586 | 2 Zn + 2 Al + PET |
| 3 | 2 Zn + 2 Al + [$C_6H_4S$]$_{10}$ | 114 | 646 | 2 Zn + 2 Al + PPS |
| 4 | 2 Zn + 2 Al + [$C_{22}H_{10}O_5N_2$]$_2$ | 82 | 478 | 2 Zn + 2 Al + PI |
| 5 | 3 Zn + 1 Al + [$C_3H_6$]$_{10}$ | 94 | 343 | 3 Zn + 1 Al + PP |
| 6 | 3 Zn + 1 Al + [$C_{10}H_8O_4$]$_5$ | 114 | 603 | 3 Zn + 1 Al + PET |
| 7 | 3 Zn + 1 Al + [$C_6H_4S$]$_{10}$ | 114 | 663 | 3 Zn + 1 Al + PPS |
| 8 | 3 Zn + 1 Al + [$C_{22}H_{10}O_5N_2$]$_2$ | 82 | 495 | 3 Zn + 1 Al + PI |

Each system was subject to the exploration of its potential energy surface. The geometries of the revealed lowest-energy stationary points representing particular interest were additionally re-optimized at the density functional theory level. This stage of research provided the amounts of gaseous by-products formed out of each simulated capacitor elemental composition and the spatial models of the solid phase remainder, which we term soot samples.

Figure 1 shows the global minimum configuration of the 2 Zn + 2 Al + PP system after the re-optimization via plane-wave KSDFT. The soot obtained from the PP polymer does not contain highly organized structures. The gas phase is represented by 17 $H_2$ molecules. In the solid phase, the chain Al…Zn…H…Al appears. The distance between aluminum and zinc atoms amounts to 273 pm. The distance between zinc and hydrogen atoms amounts to 179 pm. The distance between hydrogen and aluminum atoms amounts to 173 pm. All the above interatomic distances exceed the respective covalent radii. Thus, no covalent bonding emerges. However, the recorded interatomic distances are also smaller than the sums of the respective van der Waals radii. This observation indicates strong pairwise electrostatic attractions. Another zinc atom in this configuration appears to be built into the non-covalently bonded chain …C…Zn…C…, in which Zn is electrophilic and carbon atoms are slightly electrophilic.

Figure 2 represents the global minimum configuration of the 3 Zn + 1 Al + PP system after re-optimization via plane-wave DFT. The gas phase is represented by 15 $H_2$ molecules and two $C_2H_2$ molecules. The resulting cyclic hydrocarbon structure, including the phenyl ring, is highlighted. The chain Zn…H…Zn forms. The distances are 166 pm and 165 pm, respectively. These are larger than expected covalent bonds for zinc hydride, yet less than the sum of the van der Waals atomic radii. Hence, the electrostatic attraction exists. Aluminum gives rise to the C…Al…H chains, in which Al is electron-deficient.

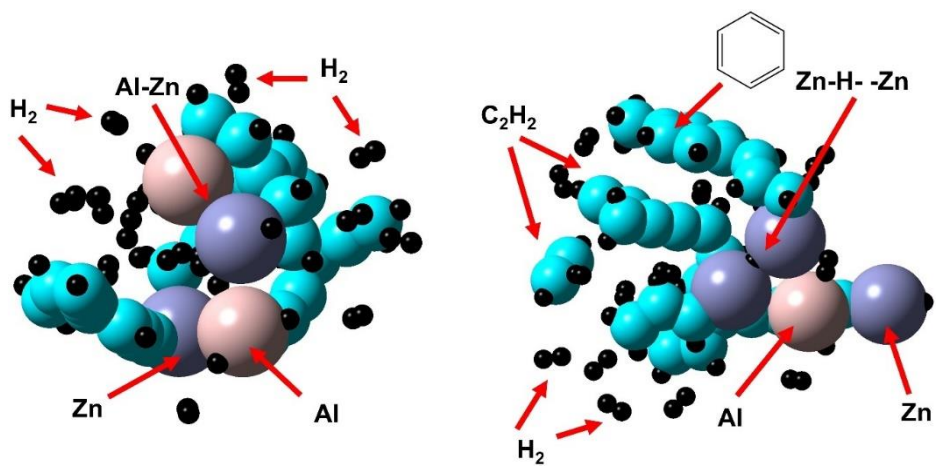

Figure 1. The global minimum configurations of the (left) [2 Zn + 2 Al + PP] system and (right) [3 Zn + 1 Al + PP] system. All depicted geometries have been obtained via hybrid density functional theory. The carbon atoms are cyan, the zinc atoms are violet, the aluminum atoms are pink, and the hydrogen atoms are black.

In the PET-containing soot sample, nine $H_2$ molecules, one CO molecule, two $H_2O$ molecules, and a single $C_2H_2$ molecule coexist in the global minimum configuration of the 2 Zn + 2 Al + PET system. Figure 8 represents the global minimum configuration of the 3 Zn + 1 Al + PET system. The gas phase is represented by nine $H_2$, three CO, and two $H_2O$ molecules. Specific ring structures with aromaticity can be observed. Zinc incorporates into the chains like

…C…Zn…O… In turn, aluminum interacts electrostatically with hydrocarbon chains. In this type of interatomic coupling, aluminum acts as Lewis acid.

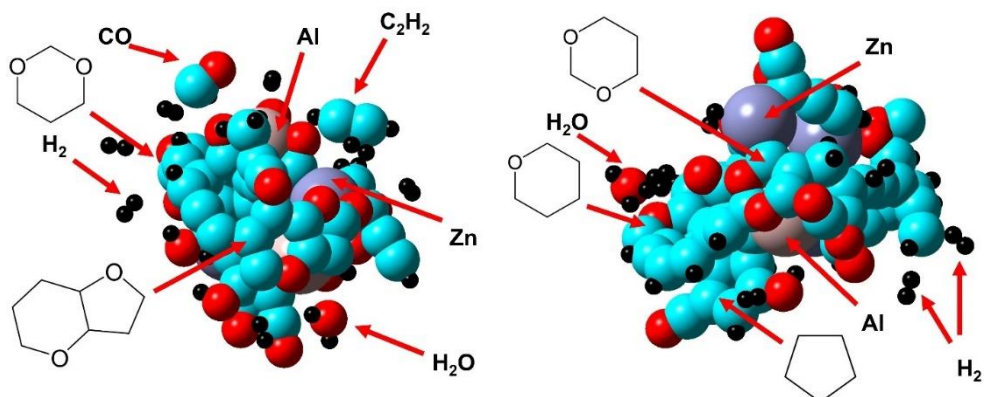

Figure 2. The global minimum configurations of the (left) [2 Zn + 2 Al + PET] system and (right) [3 Zn + 1 Al + PET] system. All depicted geometries have been obtained via hybrid density functional theory. The carbon atoms are cyan, the oxygen atoms are red, the zinc atoms are violet, the aluminum atoms are pink, and the hydrogen atoms are black.

Figure 9 shows the decomposition product of PPS containing 2 Zn and 2 Al atoms. We opted to use two atoms of each metallic element to investigate their possible interactions with one another in the soot. The formation of complex ring structures occurs. These structures include sulfur bridges, -S-S-, which is a typical chemical behavior for sulfur. Zinc atoms strongly interact with each other. The Zn…Zn distance is 255 pm, which corresponds to a strong Coulombic coupling. The distance between the zinc atom and the nearest sulfur atom amounts to 258 pm. Aluminum forms bonds with either sulfur or carbon. The Al-S distances vary between 220-230 pm in various patterns and stationary states, which corresponds to a decent covalent bonding. On the contrary, the distances between Al and C range between 196 and 210 pm, which corresponds to a strong electrostatic attraction, yet not chemical bonding. Five $H_2$ molecules and a single acetylene molecule emerge. The hydrogen sulfide gas is additionally present in some higher-energy structures. Hence, the $H_2S$ represents one of the probable products of PPS polymer pyrolysis in the capacitor.

Figure 9 illustrates the soot samples obtained in the 3 Zn + 1 Al + PPS system. Nine $H_2$ molecules form the gas phase. Complex ring structures have not been detected in this chemical composition. Aluminum incorporates into the -S-Al…Zn chain. The distance between sulfur and aluminum amounts to 227 pm. This length corresponds to the covalent bond. Because of a limited system size, we were unable to monitor the formation of the $Al_3S_2$ species. Being an inorganic salt, this structure is not expected to conduct electricity unless dissolved. The distance between aluminum and zinc is 247 pm. Zinc forms the chain Zn-H…Zn-H resembling zinc hydride. The distances between zinc and hydrogen atoms are 202, 169, and 162 pm. Therefore, covalent bonding is mixed with electrostatic attraction in this structural pattern. Sulfur is covalently built into the carbon chains of the following types, -C-S-C- and -C-S-H. The formation of S=S bridges is quite possible. The distance between sulfur atoms in such a chain equals 209 pm.

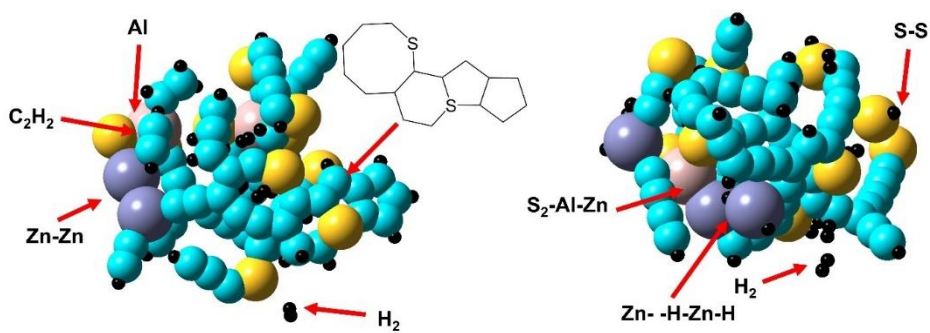

Figure 3. The global minimum configurations of the (left) [2 Zn + 2 Al + PPS] system and (right) [3 Zn + 1 Al + PPS] system. All depicted geometries have been obtained via hybrid density functional theory. The carbon atoms are cyan, the sulfur atoms are yellow, the zinc atoms are violet, the aluminum atoms are pink, and the hydrogen atoms are black.

The PI dielectric gives rise to four $H_2$ molecules and a single CO molecule in the global energy configuration of the PI-based soot with an equimolar Zn/Al ratio (Figure 10). The Zn-Al-H…Zn chain emerges. Herewith, the zinc…aluminum distance is 247 pm. The aluminum…hydrogen distance is 172 pm, and the zinc…hydrogen distance is 184 pm. The soot sample 3 Zn + 1Al + PI produces three $H_2$ molecules and five CO molecules among the gaseous by-products. Aluminum forms strong interactions with zinc, carbon, and hydrogen. The Al…Zn distance is 254 pm, the Al…C distance is 199 pm, and the Al…H distance is 160 pm. Zinc is bonded within chains of the Zn-H…Zn type. The distances between zinc and hydrogen amount to 171 and 180 pm. Zinc also forms bonds with carbon, 198 pm, and oxygen, 207 pm.

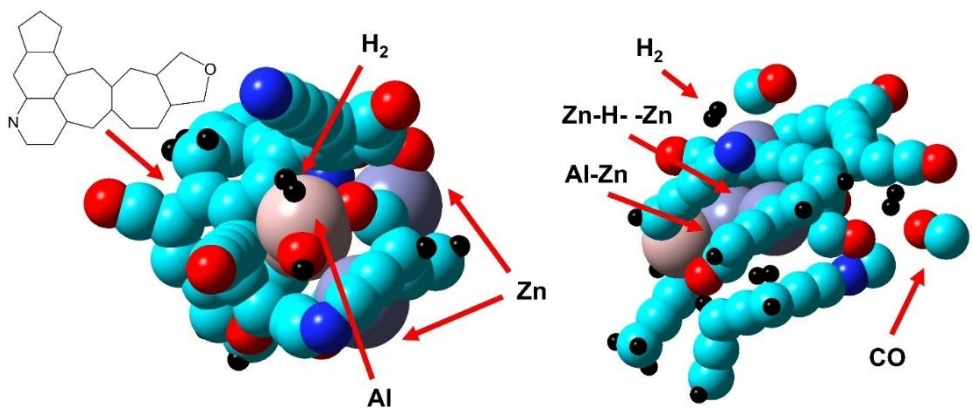

Figure 4. The global minimum configurations of the (left) [2 Zn + 2 Al + PI] system and (right) [3 Zn + 1 Al + PI] system. All depicted geometries have been obtained via hybrid density functional theory. The carbon atoms are cyan, the oxygen atoms are red, the nitrogen atoms are blue, the zinc atoms are violet, the aluminum atoms are pink, and the hydrogen atoms are black.

The global minimum atomistic configurations for each simulated composition were put in the periodic boxes, whose vectors were further optimized along with atomic coordinated at the PBEPBE pure KSDFT level as documented in the methodology section. The gases were removed from the global minimum configurations before computing electrical properties. This is because gaseous molecules do not conduct electricity.

We notice that the formed electrode atoms interact with one another in the soot. Nonetheless, they do not produce metallic clusters thanks to simultaneously coupling to the former insulator's atoms. We perceive the chemical activity of the electrode atoms as a positive parameter favoring capacitor self-healing. In this context, zinc and aluminum are more desirable electrode materials as compared to inert materials like copper, silver, or gold, despite the somewhat higher electrical conductivity of the latter. There are also other considerations dictating the choice of the electrode material in real-world dielectric capacitor technology.

Figure 5 depicts the electrical conductivities of the major soot samples as simulated at 20°C. All systems exhibit conductivities of the same order of magnitude despite substantial structural differences. The global minimum atomistic configurations for each chemical composition, reoptimized to include wave function periodicity using plane-wave KSDFT, were used. The soot sample obtained out of PPS polymer and the electrode metal atoms, Zn and Al, exhibit the highest electrical conductivity. It is threefold higher than that of PP and nearly twofold higher than that of PET. These results indicate that all investigated capacitor designs fall into the group of strong semiconductors with their estimated $\sim 10^3$ S/m conductivities. While the structure of the soot influences the conductivity, its effect is surprisingly modest. Note that the materials exhibiting conductivities of over $10^4$ S/m are formally considered to be conductors. In the meantime, metals like zinc, aluminum, copper, and others exhibit much higher conductivities than $10^4$ S/m. Consequently, the formed soot is significantly less conductive compared to electrodes and significantly more conductive compared to the insulators.

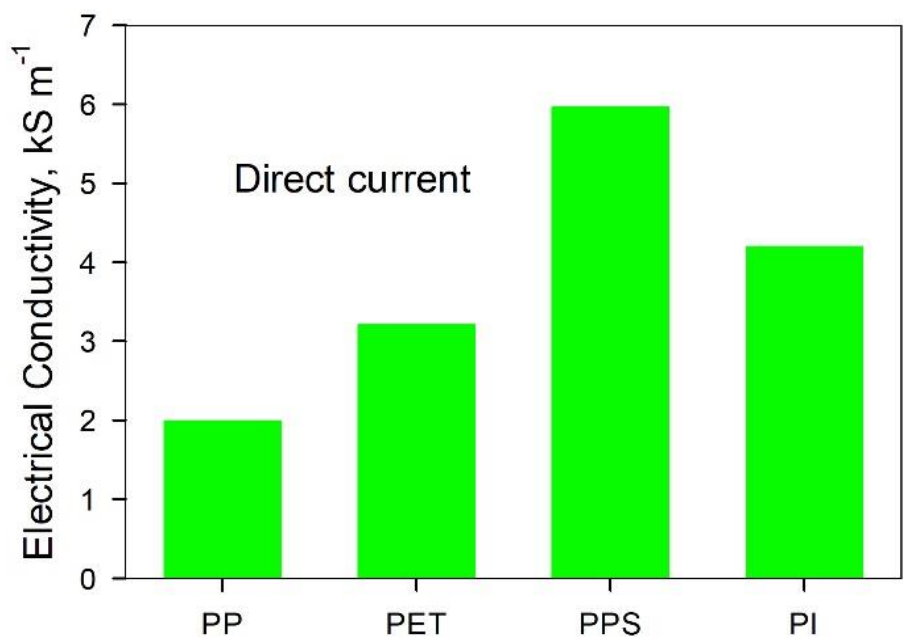

Figure 5. The electrical conductivities of the soot samples, consisting of two zinc atoms, two aluminum atoms, and four polymers at direct current at 20 °C.

Tables 2-4 summarize the parameters of the solid phase, herein coined soot, and the gas phase depending on the chemical compositions of the simulated systems. Dihydrogen was detected as an omnipresent by-product in all systems. This observation is central to capacitor self-healing.

Among other considerations, it means that maximizing the molar fraction of hydrogen atoms in the employed dielectric polymer is not enough to attain an ideal self-healing. It would have been more efficient to have these $H_2$ and $H^*$ react with the carbon atoms of the soot and thus decrease the degree of their unsaturation. The compounds possessing dangling bonds cannot minimize their potential energies and, therefore, contain high-energy electrons that are able to conduct electricity.

Table 2. The mass and molar fractions of gases in the systems, containing two zinc and two aluminum atoms with destroyed polymers. The results have been obtained from the obtained global minimum structures of the soot. The error bars do not apply to thip typer of calculation.

| Polymer | Gas species | Mass fraction | Molar fraction |
|---|---|---|---|
| PP | $17 \times H_2$ | 5.7 | 34 |
| PET | $9 \times H_2, 2 \times H_2O, 1 \times CO, 1 \times C_2H_2$ | 9.4 | 26 |
| PPS | $5 \times H_2, 1 \times C_2H_2$ | 2.8 | 15 |
| PI | $4 \times H_2, 1 \times CO$ | 3.8 | 12 |

The different proportion of the zinc and aluminum chemical elements impacts the composition of the gas phase somewhat. Since the metal atoms both covalently and non-covalently interact with the non-metallic elements, the fine effect must be expected. For instance, the valence electronic structure of Al implies stronger non-covalent interactions with electronegative atoms like oxygen, nitrogen, and sulfur. Table 2 reports the gases formed in the 2 Zn/2 Al system, whereas Table 3 reports the gases formed in the 3 Zn/1 Al system. In the PI-based composition, zinc atoms foster the formation of carbon monoxide molecules. In turn, the Al-richer PP-based system favors the dihydrogen gas. Although the effect of metal atoms can be detected, it does not change the polymer rating based on the produced volatile polymers.

Table 3. The mass and molar fractions of gases in the systems, containing three zinc atoms and one aluminum atom with destroyed polymers. The results have been obtained from the obtained global minimum structures of the soot. The error bars do not apply to thip typer of calculation.

| Polymer | Gas species | Mass fraction | Molar fraction |
|---|---|---|---|
| PP | $15 \times H_2, 2 \times C_2H_2$ | 13 | 40 |
| PET | 9×$H_2$, 2×$H_2O$, 3×CO | 13 | 26 |
| PPS | 9×$H_2$ | 1.4 | 16 |
| PI | 3×$H_2$, 5×CO | 15 | 20 |

The mass fraction of the solid phase constitutes over 90% of the electrical breakdown products in all simulated systems. First, this happens because the heavy elements do not form gaseous products. Second, the number of chemical elements that can form gases is limited in the simulated capacitor compositions. Third, the numerous unsaturated bonds, like those in carbon atoms, thermodynamically undermine the formation of separate chemical species. Previously, we have already unveiled that the enhanced fraction of the hydrogen atoms in a dielectric polymer favors its self-healing performance.

Table 4. The mass and molar fractions of the soot samples, containing two zinc and two aluminum atoms with destroyed polymers. The results have been obtained from the obtained global minimum structures of the soot. The error bars do not apply to thip typer of calculation.

| Polymer | Soot species formula | Mass fraction | Molar fraction |
|---|---|---|---|
| PP | $Zn_2Al_2C_{30}H_{26}$ | 94 | 66 |
| PET | $Zn_2Al_2C_{47}H_{16}O_{17}$ | 91 | 74 |
| PPS | $Zn_2Al_2C_{58}H_{28}S_{10}$ | 97 | 85 |
| PI | $Zn_2Al_2C_{43}H_{16}O_{10}N_4$ | 96 | 88 |

The fraction is the soot is believed to be very important for capacitor self-healing. The size of the soot sample must be made small enough to exclude its steric possibility to short-circuit the electrodes. In order to decrease the mass and volume fractions of the semiconducting soot, specific engineering measures must be taken. Various additives may change the ratios of the chemical elements to allow for specific stoichiometries of the evolving volatile by-products. On a related note, the introduction of peculiar atoms and supraatomic structures must be expected to deteriorate the conductivity of the solid phase by decreasing its local symmetries.

The primary challenge is to vaporize carbon atoms since they constitute the major fraction of the soot. Carbon forms volatile molecules with a variety of other atoms, the most obvious of them being hydrogen, oxygen, and fluorine. In the absence of competing pathways, carbon can also bind nitrogen to produce dicyane. As we revealed herein, the presence of unsaturated carbonaceous species slows the formation of volatile hydrocarbons down. This is because the carbon-carbon multiple covalent bonds in the soot are more thermodynamically beneficial compared to bonding in small hydrocarbon molecules. Additional efforts are urged to understand the most suitable reaction pathways to destroy the semiconducting soot samples, at least, to a certain extent.

**Validity of Results and Their Limitations**

The reported results are based on a few computational methods, each solving a specific task. First, a PM7-based configurational search was performed and produced 300 stationary points covering a diversity of over 1,000 kJ/mol in potential energies. Among these minimum-point geometries, a few lowest-energy ones were selected. PM7 may be in some contexts criticized for providing insufficient accuracy in the cases of high-energy geometries. These high-energy geometries are not used directly in our analysis. Yet, they may provide indirect hints regarding some insightful structural patterns to guide further research efforts. In the meantime, PM7 is known to suggest very trustworthy geometrical and thermodynamic properties even of very exotic chemicals.[30-31] In all cases containing questionable structural patterns based on conventional chemical wisdom, the PBEPBE KSDFT was involved to verify the predictions. This work reports only DFT-based results in Figures 1-4 and Tables 2-4. PM7 represents the only adequate model Hamiltonian for the performed endeavor because of its aggregate computational cost.

The plane-wave PBEPBE functional-based KSDFT was used to determine the electrical conductivities of the soot samples. The global minimum atomistic configurations obtained from PM7 were used. We assume that the global minimum atomistic configurations based on PM7 and PBEPBE are similar enough. While suggested global minima can obviously be somewhat different in each particular chemical composition, they remain to represent very low energy geometries according to any trustworthy model chemistry. Before the productive simulations, the coordinates of the atoms and cell vectors were reoptimized to correspond to a new, KSDFT, model chemistry. The pure KSDFT functional may be inaccurate for band gaps, Fermi energy, and related properties, such as electrical conductivity. Yet, it is still the most widely used approach in state-of-the-art computational physics to study semiconductors. On a related note, the order of magnitude of electrical conductivities that we derived looks to fit the expected region of values, $10^{-4}$-$10^{-3}$ S/m. Unfortunately, we are not aware of any relevant experimental work to directly compare the computed conductivities of the capacitor soot with.

The sizes of the simulated systems may be the most severe limitation in the current simulations. Because of a tiny model size, the formation of certain large structures may be inaccessible. While we did not see such giant formations in the simulated low-energy conformations, there may be some drawbacks associated with that. Furthermore, large metallic clusters cannot principally form with the present simulation setup. Yet, we did not observe the implications supporting such structures since both zinc and aluminum tend to bind non-metals as long as the latter are present in the vicinity. Unfortunately, the size limitation is unavoidable given the methods employed. Obviously, the soot samples in the real world are much larger, ~$10^{-6}$ m, than the constructed models, ~$10^{-9}$ m. We strove to keep the sizes of the systems similar so that the same restrictions apply to all simulated chemical compositions. The actual sizes were dictated by the expected computational costs and self-consistent-field procedure efficient convergence. Note that the wave funvtion convergence speed is inversely proportional to the system size in the units of valence electrons.

Since the reliability of the proposed analysis completely relies on the soot geometries, we performed numerous iterations, 200-300, to find very thermodynamically stable structural patterns at all stages of these composite simulations. As of today, we are not informed of any assumptions or approximations that could change our hereby reported predictions qualitatively.

**Conclusions and Final Considerations**

In the present work, we reported the chemical products of the electrical breakdown in metalized-film dielectric capacitors, consisting of various electrodes and polymeric films. We discussed the chemical identity of the decomposition by-products and analyzed their impact on the efficacies of capacitor self-healing. We confirmed that PP and PET foster self-healing to the largest currently known extent. Not only do these dielectric films lose the biggest fractions of the volumes of their solid counterparts also exhibit the lowest electrical conductivities out of the investigated dataset. The performances of PI and PPS are less encouraging in the context of self-healing. This conclusion agrees with our laboratory observations on the lifespans of the respective capacitor compositions (PP, PET, PPS, and PI).

While the electrical conductivities obtained for the samples of various chemical compositions, indeed, differ by up to three times, it remains insufficient to consider them principally different. All samples exhibit the behavior of semiconductors, that is, they can occasionally conduct the current and cause short circuits. The size of the soot sample also matters greatly. If the emerged semiconductor lacks a dimension to form a bridge, its conductivity per se cannot harm the capacitor. One of the most important conclusions of the last findings is, that one can rate the capacitor designs in terms of the emerging gaseous by-products, ignoring the precise value of the conductivity.

The semiconductor behavior of the soot can be directly linked to the presence of unsaturated carbon-based fragments. Indeed, it is well-known that carbon, silicon, and heavier elements in Group 14 exhibit semiconducting and even conductive behaviors depending on their state of hybridization. Mind long carbon nanotubes with specific topologies as a broadly known example of electrical conductivity by non-metallic elements. Unfortunately, the hydrogen atoms readily form dihydrogen molecules after the electrical breakdown instead of saturating potentially conductive aromatic carbon rings.

Carbon-rich soot structures are probably unavoidable as long as carbon-based polymeric insulators are employed. Since it is not possible to produce insulating soot, other options must be trialed. For instance, mineral oils, which are employed to soak the insulators for better electrode/insulator contact,[12] can be chemically adjusted to contain the elements quenching the conductivity of the soot. Research efforts are urged to reliably pick up such elements. Ozonation of the insulator means to increase the content of osygen atoms, at least, at the surface of the insulator. In combination of hydrogen atoms, oxygen must produce water at the temperature of the electrical breakdown. The ozonation may also be interesting to change the chemical composition of the polymer, therefore, adjusting the elemental ratios. In general, the introduction of unusual atoms in the structure of the soot is deemed to decrease the conductivity because they lower the level of symmetry and inevitably perturb regular electronic structures as a result.

**Credit Author Statement**

Author 1: Conceptualization; Methodology Development; Software development; Validation; Formal analysis; Investigation; Resources; Data Curation; Writing - Original Draft; Writing - Review & Editing; Visualization Preparation; Supervision; Project administration; Funding acquisition.

Author 2: Visualization Preparation.

**Conflict of Interest**

The authors hereby declare no financial interests and professional connections that might bias the interpretations of the obtained results.

**Data Availability**

All data supporting this article has been included in the publication.

**Acknowledgments**

The research was funded by the Ministry of Science and Higher Education of the Russian Federation under the strategic academic leadership program "Priority 2030" (Agreement 075-15-2024-201 dated 06.02.2024). The results of the work were obtained using computational resources of Peter the Great Saint-Petersburg Polytechnic University Supercomputing Center (www.spbstu.ru). We thank Dr. Eugeny Petukhov for his extensive support and interesting technical discussions. V.V.C. is an invited foreign professor at YSU in Yerevan.